   \documentclass[12pt]{article}
   \usepackage{latexsym}
\else\oddsidemargin25mm\evensidemargin25mm\marginparwidth25mm\fi%
\def\theequation{\arabic{section}.\arabic{equation}}

\renewcommand{\theequation}{\thesection.\arabic{equation}}
\begin{document}
%%%%%%%%%%%%%%%%%%%%%%%%%%%%%%%%%%%%%%%%%%
%%%%%%%%%%%%%%%%%%%%%%%%%%%%%%%%%%%%%%%%%%
% %%%%%%%%%%%%%%%%%%%%%%%%%%%%%%%%%%%%%%%%%
%%%%%% ANNAMACRO %%%%%%%%%%%%%%%%%%%%%%%%%%
%%%%%%%%%%%%%%%%%%%%%%%%%%%%%%%%%%%%%%%%%%%
%%%%%%%%%%%%%%%%%%%%%%%%%%%%%%%%%%%%%%%%%%%
\makeatletter \@addtoreset{equation}{section} \makeatother
\renewcommand{\theequation}{\thesection.\arabic{equation}}
\newcommand{\ft}[2]{{\textstyle\frac{#1}{#2}}}
\newcommand{\QED}{{\hspace*{\fill}\rule{2mm}{2mm}\linebreak}}
\def\dop{{\rm d}\hskip -1pt}
\def\bfone{\relax{\rm 1\kern-.35em 1}}
\def\bfzero{\relax{\rm I\kern-.18em 0}}
\def\inbar{\vrule height1.5ex width.4pt depth0pt}
\def\IC{\relax\,\hbox{$\inbar\kern-.3em{\rm C}$}}
\def\ID{\relax{\rm I\kern-.18em D}}
\def\IF{\relax{\rm I\kern-.18em F}}
\def\IK{\relax{\rm I\kern-.18em K}}
\def\IH{\relax{\rm I\kern-.18em H}}
\def\II{\relax{\rm I\kern-.17em I}}
\def\IN{\relax{\rm I\kern-.18em N}}
\def\IP{\relax{\rm I\kern-.18em P}}
\def\IQ{\relax\,\hbox{$\inbar\kern-.3em{\rm Q}$}}
\def\IR{\relax{\rm I\kern-.18em R}}
\def\IG{\relax\,\hbox{$\inbar\kern-.3em{\rm G}$}}
\font\cmss=cmss10 \font\cmsss=cmss10 at 7pt
\def\ZZ{\relax\ifmmode\mathchoice
{\hbox{\cmss Z\kern-.4em Z}}{\hbox{\cmss Z\kern-.4em Z}}
{\lower.9pt\hbox{\cmsss Z\kern-.4em Z}} {\lower1.2pt\hbox{\cmsss
Z\kern .4em Z}}\else{\cmss Z\kern-.4em Z}\fi}
\def\a{\alpha} \def\b{\beta} \def\d{\delta}
\def\e{\epsilon} \def\c{\gamma}
\def\G{\Gamma} \def\l{\lambda}
\def\L{\Lambda} \def\s{\sigma}
\def\cA{{\cal A}} \def\cB{{\cal B}}
\def\cC{{\cal C}} \def\cD{{\cal D}}
    \def\cF{{\cal F}} \def\cG{{\cal G}}
\def\cH{{\cal H}} \def\cI{{\cal I}}
\def\cJ{{\cal J}} \def\cK{{\cal K}}
\def\cL{{\cal L}} \def\cM{{\cal M}}
\def\cN{{\cal N}} \def\cO{{\cal O}}
\def\cP{{\cal P}} \def\cQ{{\cal Q}}
\def\cR{{\cal R}} \def\cV{{\cal V}}\def\cW{{\cal W}}
%
%
%%%%%%%%%%%%%%%%%%%%%%%%%%%%%%%%%%%%%%%%%%%%%%%%%%%%%%%%%%%%%%
%%%%% misc macros %%%%%
%%%%%%%%%%%%%%%%%%%%%%%%%%%%%%%%%%%%%%%%%%%%%%%%%%%%%%%%%%%%%%
%
\def\crr{\crcr\noalign{\vskip {8.3333pt}}}
\def\tilde{\widetilde}
\def\bar{\overline}
\def\us#1{\underline{#1}}
\let\shat=\hat
\def\hat{\widehat}
\def\hyp{\vrule height 2.3pt width 2.5pt depth -1.5pt}
\def\square{\mbox{.08}{.08}}
\def\Coeff#1#2{{#1\over #2}}
\def\Coe#1.#2.{{#1\over #2}}
\def\coeff#1#2{\relax{\textstyle {#1 \over #2}}\displaystyle}
\def\coe#1.#2.{\relax{\textstyle {#1 \over #2}}\displaystyle}
\def\half{{1 \over 2}}
\def\shalf{\relax{\textstyle {1 \over 2}}\displaystyle}
\def\dag#1{#1\!\!\!/\,\,\,}
\def\to{\rightarrow}
\def\notin{\hbox{{$\in$}\kern-.51em\hbox{/}}}
\def\shdot{\!\cdot\!}
\def\ket#1{\,\big|\,#1\,\big>\,}
\def\bra#1{\,\big<\,#1\,\big|\,}
\def\equaltop#1{\mathrel{\mathop=^{#1}}}
\def\Trbel#1{\mathop{{\rm Tr}}_{#1}}
\def\inserteq#1{\noalign{\vskip-.2truecm\hbox{#1\hfil}
\vskip-.2cm}}
\def\attac#1{\Bigl\vert
{\phantom{X}\atop{{\rm\scriptstyle #1}}\phantom{X}}}
\def\exx#1{e^{{\displaystyle #1}}}
\def\del{\partial}
\def\delbar{\bar\partial}
\def\nex#1{$N\!=\!#1$}
\def\dex#1{$d\!=\!#1$}
\def\cex#1{$c\!=\!#1$}
\def\eg{{\it e.g.}} \def\ie{{\it i.e.}}
%\catcode`\@=12
%%%%%%%%%%%%%%%%%%%%%%%%%%%%%%%%%%%%%%%%%%%%%%%%%%%%%%%%%%%%%%
%%%%%%%%%%%%%%%%%%%%%%%%%%%%%%%%%%%%%%%%%%%%%%%%%%%%%%%%%%%%
%\draft
%%%%%%%%%%%% macros and references %%%%%%%%%%%%%%%%%%%%%%%%%
%
\def\cS{{\cal K}}
\def\IE{\relax{{\rm I\kern-.18em E}}}
\def\cE{{\cal E}}
\def\rt{{\cR^{(3)}}}
\def\IGam{\relax{{\rm I}\kern-.18em \Gamma}}
\def\IGa{\IA}
\def\LG{Lan\-dau-Ginz\-burg\ }
\def\cV{{\cal V}}
\def\Rt{{\cal R}^{(3)}}
\def\wabc{W_{abc}}
\def\WABC{W_{\a\b\c}}
\def\W{{\cal W}}
\def\tft#1{\langle\langle\,#1\,\rangle\rangle}
\def\IA{\relax{\hbox{{\rm A}\kern-.82em {\rm A}}}}
\let\picfuc=\fp
\def\hata{{\shat\a}}
\def\hatb{{\shat\b}}
\def\hatA{{\shat A}}
\def\hatB{{\shat B}}
\def\bv{{\bf V}}
\def\spg{special geometry}
\def\sc{SCFT}
\def\leel{low energy effective Lagrangian}
\def\pf{Picard--Fuchs}
\def\pfS{Picard--Fuchs system}
\def\el{effective Lagrangian}
\def\Fb{\overline{F}}
\def\nablab{\overline{\nabla}}
\def\Ub{\overline{U}}
\def\Db{\overline{D}}
\def\zb{\overline{z}}
\def\eb{\overline{e}}
\def\fb{\overline{f}}
\def\tb{\overline{t}}
\def\Xb{\overline{X}}
\def\Vb{\overline{V}}
\def\Cb{\overline{C}}
\def\Sb{\overline{S}}
\def\delb{\overline{\del}}
\def\Gammab{\overline{\Gamma}}
\def\Ab{\overline{A}}
\def\Anh{A^{\rm nh}}
\def\alphab{\bar{\alpha}}
\def\cy{Calabi--Yau}
\def\cabg{C_{\alpha\beta\gamma}}
\def\B{\Sigma}
\def\Bh{\hat \Sigma}
\def\Kh{\hat{K}}
\def\Knh{{\cal K}}
\def\A{\Lambda}
\def\Ah{\hat \Lambda}
\def\R{\hat{R}}
\def\V{{V}}
\def\T{T}
\def\Gammah{\hat{\Gamma}}
\def\twot{$(2,2)$}
\def\K{K\"ahler}
\def\rat{({\theta_2 \over \theta_1})}
\def\lv{{\bf \omega}}
\def\w{w}
\def\CP{C\!P}
\def\o#1#2{{{#1}\over{#2}}}
%%%%%%%%%%%%%%%%%%%%%%%%%%%%%%%%%%%%%%%%%%%%%%%%%%%%%%%
% FINE DELLA ANNAMACRO.TEX %%%%%%%%%%%%%%%%%%%%%%%%%%%%
%%%%%%%%%%%%%%%%%%%%%%%%%%%%%%%%%%%%%%%%%%%%%%%%%%%%%%%
%%%%%%%%%%%%%%%%%%%%%%%%%%%%%%%%%%%%%%%%%%%%%%%%%%%%%%%
% Paolo Macro %%%%%%%%%%%%%%%%%%%%%%%%%%%%%%%%%%%%%%%%%
%%%%%%%%%%%%%%%%%%%%%%%%%%%%%%%%%%%%%%%%%%%%%%%%%%%%%%%
\newcommand{\be}{\begin{equation}}
\newcommand{\ee}{\end{equation}}
\newcommand{\ba}{\begin{eqnarray}}
\newcommand{\ea}{\end{eqnarray}}
\newtheorem{definizione}{Definition}[section]
\newcommand{\bd}{\begin{definizione}}
\newcommand{\ed}{\end{definizione}}
\newtheorem{teorema}{Theorem}[section]
\newcommand{\bth}{\begin{teorema}}
\newcommand{\eth}{\end{teorema}}
\newtheorem{lemma}{Lemma}[section]
\newcommand{\blem}{\begin{lemma}}
\newcommand{\elem}{\end{lemma}}
\newcommand{\brr}{\begin{array}}
\newcommand{\err}{\end{array}}
\newcommand{\nn}{\nonumber}
\newtheorem{corollario}{Corollary}[section]
\newcommand{\bcorol}{\begin{corollario}}
\newcommand{\ecorol}{\end{corollario}}
%%%%%%%%%%%%%%%%%%%%%%%%%%%%%%%%%%%%%%%%%%%%%%
\def\twomat#1#2#3#4{\left(\begin{array}{cc}
 {#1}&{#2}\\ {#3}&{#4}\\
\end{array}
\right)}
\def\twovec#1#2{\left(\begin{array}{c}
{#1}\\ {#2}\\
\end{array}
\right)}
%%%%%%%%%%%%%%%%%%%%%%%%%%%%%%%%%%%%%%%%%%%%%%%%%%%%
%%%%%%%%%%%%%%%%%%%%%%%%%%%%%%%%%%%%%%%%%%%%%%%%%%%%%%%%%%%%%
%% This File is the titlepage %%%%%%%%%%%%%%%%%%%%%%%%%%%%%%%%
%%%%%%%%%%%%%%%%%%%%%%%%%%%%%%%%%%%%%%%%%%%%%%%%%%%%%%%%%%%%%
%\begin{titlepage}
%\hskip 5.5cm
%\vbox{
%\hbox{CERN-TH/2000-167}
%}
\hskip 7.5cm
\vbox{\hbox{CERN-TH/2000-255}\hbox{hep-th/0008xx}\hbox{August,
2000}} %\vfill
 \vskip 2cm
\begin{center}
{\LARGE { $F(4)$ Supergravity and $5D$ Superconformal Field
Theories}\footnote{The work of R. D'Auria and S. Vaul\'a has been
supported by EEC under TMR contract ERBFMRX-CT96-0045, the work
of S. Ferrara has been supported by the EEC TMR programme
ERBFMRX-CT96-0045 (Laboratori Nazionali di Frascati, INFN) and by
DOE grant DE-FG03-91ER40662, Task C.}}\\
%\vfill
\vskip 1.5cm
  {\bf Riccardo D'Auria$^1$, Sergio Ferrara$^2$ and
Silvia Vaul\'a$^3$ } \\
%\vfill
\vskip 0.5cm {\small $^1$ Dipartimento di Fisica, Politecnico di
Torino,\\
 Corso Duca degli Abruzzi 24, I-10129 Torino\\
and Istituto Nazionale di Fisica Nucleare (INFN) - Sezione di
Torino, Italy}\\
{\small $^2$ CERN Theoretical Division, CH 1211 Geneva 23, Switzerland\\
and Istituto Nazionale di Fisica Nucleare, Laboratori Nazionali
di Frascati, Italy
}\\ {\small $^3$ Dipartimento di Fisica Teorica,
Universit\'a di Torino, via P. Giuria 1, I-10125 Torino}
\vspace{6pt}
\end{center}
\vskip 3cm {\sl Proceedings of the Gursey Memorial Conference
II,"M-theory and Dualities", June 19-23 2000, Istanbul, Turkey.}
%\vfill
%\begin{center} {\bf Abstract}
%\end{center}
%\end{titlepage}
%%%%%%%%%%%%%%%%%%%%%%%%%%%%%%%%%%%%%%%%%%%%%%%%%%%%%%%%%%%%%%%%%%%%%%
%%%%%%%%%%%%               INTRODUZIONE                %%%%%%%%%%%%%%%%
%%%%%%%%%%%%%%%%%%%%%%%%%%%%%%%%%%%%%%%%%%%%%%%%%%%%%%%%%%%%%%%%%%%%%%%
\section{Introduction}

We report on a recent investigation \cite{nostro}in which $F(4)$
supergravity \cite{rom}, the gauge theory of the exceptional
six-dimensional Anti-de Sitter superalgebra \cite{nahm,bagu}, is
coupled  to an arbitrary number of vector multiplets whose scalar
components parametrize the quaternionic manifold
$SO(4,n)/SO(4)\times SO(n)$. By gauging the compact subgroup
$SU(2)_d \otimes \cG$, where $SU(2)_d$ is the diagonal subgroup
of $SO(4)\simeq SU(2)_L\otimes SU(2)_R$ (the$R$-symmetry group of
six-dimensional Poincar\'e supergravity) and $\cG$ is a compact
group such that $dim\cG = n$, we obtain the scalar potential
which, besides the gauge coupling constants, also depends in non
trivial way on the parameter $m$ associated to a massive 2-form
$B_{\mu\nu}$ of the gravitational multiplet. The potential admits
an $AdS$ background for $g=3m$, as the pure $F(4)$-supergravity.
We compute the scalar squared masses (which are all
negative)which are found to match the results dictated by
$AdS_6/CFT_5$ correspondence \cite{malda,rass}from the conformal
dimensions of boundary operators. The boundary $F(4)$
superconformal fields are realized in terms of a singleton
superfield (hypermultiplet) in harmonic superspace with flag
manifold $SU(2)/U(1)=S^2$. Finally we analyze the spectrum of
short representations in terms of  superconformal primaries and
predict general features of the K-K spectrum of massive type IIA
supergravity compactified on warped $AdS_6\otimes S^4$.

\section{A geometrical setting}
In this section we set up a suitable framework for the discussion
of the matter coupled $F(4)$ supergravity theory and its gauging.
\noindent This will allow us to set up the formalism for the
matter coupling in the next section. Actually we will just give
the essential definitions of the Bianchi identities approach in
superspace , while all the relevant results, specifically the
supersymmetry transformation laws of the fields, will be given in
the ordinary space-time formalism.

First of all it is useful to discuss the main results of ref.
\cite{rom} by a careful study in superspace of the Poincar\'e and
AdS supersymmetric vacua.
 Let us recall the content of $D=6$, $N=(1,1)$ supergravity
multiplet:
\begin{equation}
(V^{a}_{\mu}, A^{\alpha}_{\mu}, B_{\mu\nu},\,\ \psi^{A}_{\mu},\,\
\psi^{\dot{A}}_{\mu}, \chi^{A}, \chi^{\dot{A}}, e^{\sigma})
\end{equation}
\noindent where $V^a_{\mu}$ is the six dimensional vielbein,
$\psi^{A}_{\mu},\,\ \psi^{\dot{A}}_{\mu}$ are left-handed and
right-handed four- component gravitino fields respectively, $A$
and $\dot{A}$ transforming under the two factors of the
$R$-symmetry group $O(4)\simeq SU(2)_L\otimes SU(2)_R$,
$B_{\mu\nu}$ is a 2-form, $A^{\alpha}_{\mu}$ ($\alpha=0,1,2,3$),
are  vector fields, $\chi^{A}, \chi^{\dot{A}}$ are left-handed and
right-handed spin $\frac{1}{2}$ four components dilatinos, and
$e^{\sigma}$ denotes the dilaton.\\ Our notations are as follows:
$a,b,\dots=0,1,2,3,4,5$ are Lorentz flat indices in $D=6$
$\mu,\nu,\dots=0,1,2,3,4,5$ are the corresponding world indices,
  $A,\dot{A}=1,2$. Moreover our metric is
$(+,-,-,-,-,-)$.\\ We recall that the description of the spinors
of the multiplet in terms of left-handed and right-handed
projection holds only in a Poincar\'e background, while in an AdS
background the chiral projection cannot be defined and we are
bounded to use 8-dimensional pseudo-Majorana spinors. In this
case the $R$-symmetry group reduces to the $SU(2)$ subgroup of
$SU(2)_L\otimes SU(2)_R$, the $R$-symmetry group of the chiral
spinors. For our purposes, it is convenient to use from the very
beginning 8-dimensional pseudo-Majorana spinors even in a
Poincar\'e framework, since we are going to discuss in a unique
setting both Poincar\'e and $AdS$ vacua.\\ The pseudo-Majorana
condition on the gravitino 1-forms is as follows:
\begin{equation}
(\psi_A)^{\dagger}\gamma^0=\overline{(\psi_A)}=\epsilon^{AB}\psi_{B}^{\
\ t}
\end{equation}
\noindent where we have chosen the charge conjugation matrix in
six dimensions as the identity matrix (an analogous definition
six dimensions as the identity matrix (an analogous definition
holds for the dilatino fields). We use eight dimensional
antisymmetric gamma matrices, with
$\gamma^7=i\gamma^0\gamma^1\gamma^2\gamma^3\gamma^4\gamma^5$,
which implies $\gamma_7^T=-\gamma_7$ and $(\gamma_7)^2=-1$. The
indices $A,B,\dots=1,2,$ of the spinor fields $\psi_A,\,\ \chi_A$
transform in the fundamental of the diagonal subgroup $SU(2)$ of
$SU(2)_L\otimes SU(2)_R$. For a generic $SU(2)$ tensor $T$,
raising and lowering of indices are defined by
\begin{eqnarray}
&&T^{\dots A\dots}=\epsilon^{AB}\ \ T^{\dots\ \ \dots}_{\ \ B}\\
&&T_{\dots A\dots}=T_{\dots\ \ \dots}^{\ \ B}\ \ \epsilon_{BA}
\end{eqnarray}

 To study the supersymmetric vacua let us write down the Maurer-Cartan
Equations (M.C.E.) dual to the $F(4)$ Superalgebra
(anti)commutators:

\begin{eqnarray}
\label{dVx}&&\mathcal{D}V^{a}-\frac{i}{2}\ \
\overline{\psi}_{A}\gamma_{a}\psi^A =0\\
&&\mathcal{R}^{ab}+4m^{2}\ \
V^{a}V^{b}+m\overline{\psi}_{A}\gamma_{ab}\psi^A=0\\
&&dA^{r}+\frac{1}{2}\,\ g\,\ \epsilon^{rst}A_{s}A_{t}-i\,\
\overline{\psi}_{A}\psi_{B}\,\
\sigma^{rAB}=0\\
\label{dPsix}&&D\psi_{A}-im\gamma_{a}\psi_{A}V^{a}=0
\end{eqnarray}

\noindent Here $V^a,\omega^{ab},\psi_A,A^r,(r= 1,2,3)$, are
superfield 1-forms dual to the $F(4)$ supergenerators  which at
$\theta =0$ have as $dx^{\mu}$ components
\begin{equation}
V^a_{\mu}= \delta^a_{\mu},\,\ \psi_{A\mu}=A^r_{\mu}=0, \,\
\omega^{ab}_{\mu}= pure\,\ gauge.
\end{equation}
Furthermore $\mathcal{R}^{ab}\equiv
d\omega^{ab}-\omega^{ac}\land\,\ \omega_c^{\,\ b}$, $\mathcal{D}$
is the Lorentz covariant derivative, $D$ is the $SO(1,5)\otimes
SU(2)$ covariant derivative, which on spinors acts as follows:
\begin{equation}D\psi_A\equiv
d\psi_A-\frac{1}{4}\gamma_{ab}\omega^{ab}\psi_A-\frac{i}{2}\sigma_{AB}^r
A_r\psi^B
\end{equation}
\noindent Note that $\sigma^{rAB}=\epsilon^{BC}\sigma^{rA}_{\ \
C}$, where $\sigma^{rA}_{\ \ B}$\ \ ($r=1,2,3$) denote the usual
Pauli matrices, are symmetric in $A,\,\ B$.\\ Let us point out
that the $F(4)$ superalgebra, despite the presence of two
different physical parameters, the $SU(2)$ gauge coupling constant
$g$ and the inverse $AdS$ radius $m$, really depends on just one
parameter since the closure under $d$-differentiation of eq.
(\ref{dPsix})  (equivalent to the implementation of Jacobi
identities on the generators), implies $g=3m$; to recover this
result one has to use the following Fierz identity involving
3-$\psi_A$'s 1-forms:
\begin{equation}
\label{fond}\frac{1}{4}\gamma_{ab}\psi_A\overline{\psi}_B\gamma^{ab}\psi_C\epsilon^{AC}-\frac{1}{2}
\gamma_{a}\psi_A\overline{\psi}_B\gamma^{a}\psi_C\epsilon^{AC}+3\psi_C\overline{\psi}_B\psi_A\epsilon^{BC}=0
\end{equation}\\The $F(4)$ superalgebra described by equations
(\ref{dVx}) -(\ref{dPsix}) fails to describe the physical vacuum
because of the absence of the superfields 2-form $B$ and 1-form
$A^0$ whose space-time restriction coincides with the physical
fields $B_{\mu\nu}$ and $A_{\mu}^0$ appearing in the supergravity
multiplet.
 The recipe to have all the fields in
a single algebra is well known and consists in considering the
Free Differential Algebra (F.D.A.)\cite{bible} obtained from the
$F(4)$ M.C.E.'s by adding two more equations for the 2-form $B$
and for the 1-form $A^0$ (the 0-form fields $\chi_A$ and $\sigma$
do not appear in the algebra since they are set equal to zero in
the vacuum). It turns out that to have a consistent F.D.A.
involving $B$ and $A^0$ one has to add to the $F(4)$ M.C.E.'s two
more equations involving $dA^0$ and $dB$; in this way one obtains
an extension of the M.C.E's to the following F.D.A:
\begin{eqnarray}\label{dV}&&\mathcal{D}V^{a}-\frac{i}{2}\ \
\overline{\psi}_{A}\gamma_{a}\psi^A=0 \\
\label{dO}&&\mathcal{R}^{ab}+4m^{2}\ \
V^{a}V^{b}+m\overline{\psi}_{A}\gamma_{ab}\psi^A=0\\
 \label{dAr}&&dA^{r}+\frac{1}{2}\,\ g\,\
\epsilon^{rst}A_{s}A_{t}-i\,\ \overline{\psi}_{A}\psi_{B}\,\
\sigma^{rAB}=0\\
\label{dA}&&dA^0-mB-i\,\
\overline{\psi}_{A}\gamma_{7}\psi^A=0\\
\label{dB}&&dB+2\,\
\overline{\psi}_{A}\gamma_{7}\gamma_{a}\psi^AV^{a}=0\\
\label{dPsi}&&D\psi_{A}-im\gamma_{a}\psi_{A}V^{a}=0\end{eqnarray}
\noindent  Equations (\ref{dA}) and (\ref{dB}) were obtained by
imposing that they satisfy the $d$-closure together with equations
(\ref{dV}). Actually the closure of (\ref{dB}) relies on the
4-$\psi_A$'s Fierz identity
\begin{equation}
\overline{\psi}_{A}\gamma_{7}\gamma_{a}\psi_{B}\epsilon^{AB}\overline{\psi}_{C}\gamma^{a}\psi_{D}\epsilon^{CD}=0
\end{equation}
The interesting feature of the F.D.A (\ref{dV})-(\ref{dPsi}) is
the appearance of the combination $dA^0-mB$ in (\ref{dA}). That
means that the dynamical theory obtained by gauging the F.D.A.
out of the vacuum will contain the fields $A^0_{\mu}$ and
$B_{\mu\nu}$ always in the single combination
$\partial_{[\mu}A^0_{\nu]}-mB_{\mu\nu}$. At the dynamical level
this implies, as noted by Romans \cite{rom}, an Higgs phenomenon
where the 2-form $B$ "eats" the 1-form $A^0$ and acquires a non
vanishing mass $m$ \footnote{An analogous phenomenon takes place
also in $D=5$; see\cite{anna}.}.\\ In summary, we have shown that
two of the main results of \cite{rom}, namely  the existence of
an $AdS$ supersymmetric background only for $g=3m$ and  the
Higgs-type mechanism by which the field $B_{\mu\nu}$ becomes
massive acquiring longitudinal degrees of freedom in terms of the
the vector $A^0_{\mu}$, are a simple consequence of the algebraic
structure of the F.D.A. associated to the $F(4)$ supergroup
written in terms of the
vacuum-superfields.\\
It is interesting to see what happens if one or both the
parameters $g$ and $m$ are zero. Setting $m=g=0$, one reduces the
$F(4)$ Superalgebra to the $D=6\ \ N=(1,1)$ superalgebra existing
only in a Super Poincar\'e background; in this case the four-
vector $A^{\alpha}\equiv (A^0,A^r)$ transforms in the fundamental
of the $R$-symmetry group $SO(4)$ while the pseudo-Majorana
spinors $\psi_A,\chi_A$ can be decomposed in two chiral spinors
in such a way that all the resulting F.D.A. is invariant under
$SO(4)$.\\
Furthermore it is easy to see that no F.D.A  exists if either
$m=0$ , $g\neq 0$ or $m\neq 0$, $g= 0$, since the corresponding
equations in the F.D.A. do not close anymore under $d$-
differentiation. In other words the gauging of $SU(2)$, $g\neq 0$
must be necessarily accompanied by the presence of the parameter
$m$ which, as we have
 seen, makes the closure of the supersymmetric algebra consistent for $g=3m$.\\

In $D=6,\,\ N=4$ Supergravity, the only kind of matter is given
by vector multiplets, namely
\begin{equation}
(A_{\mu},\,\ \lambda_A,\,\ \phi^{\alpha})^I
\end{equation}
\noindent where $\alpha=0,1,2,3$ and the index $I$ labels an
arbitrary number $n$ of such multiplets. As it is well known the
$4n$ scalars parametrize the coset manifold $SO(4,n)/SO(4)\times
SO(n)$. Taking into account that the pure supergravity has a non
compact duality group $O(1,1)$ parametrized by $e^{\sigma}$, the
duality group of the matter coupled theory is
 \begin{equation}\label{coset}
  G/H=\frac{SO(4,n)}{SO(4)\times SO(n)}\times O(1,1)
\end{equation}
To perform the matter coupling we follow the geometrical
procedure of introducing the coset representative $L^{\Lambda}_{\
\ \Sigma}$ of the matter coset manifold, where
$\Lambda,\Sigma,\dots=0, \dots, 3+n$; decomposing the $O(4,n)$
indices with respect to $H=SO(4)\times O(n)$ we have:
\begin{equation}
L^{\Lambda}_{\ \ \Sigma}=(L^{\Lambda}_{\ \ \alpha},L^{\Lambda}_{\
\ I})
\end{equation}
\noindent where $\alpha=0,1,2,3$, $I=4,\dots ,3+n$. Furthermore,
since we are going to gauge the $SU(2)$ diagonal subgroup of
$O(4)$ as in pure Supergravity, we will also decompose
$L^{\Lambda}_{\ \ \alpha}$ as
\begin{equation}
L^{\Lambda}_{\ \ \alpha}=(L^{\Lambda}_{\ \ 0}, L^{\Lambda}_{\ \
r})
\end{equation}
The $4+n$ gravitational and matter vectors will now transform in
the fundamental of $SO(4,n)$ so that the superspace vector
curvatures will be now labeled by the index $\Lambda$:
$F^{\Lambda} \equiv (F^0,F^r,F^I)$. Furthermore the covariant
derivatives acting on the spinor fields will now contain also the
composite connections of the $SO(4,n)$ duality group. Let us
introduce the left-invariant 1-form of $SO(4,n)$
\begin{equation}
\Omega^{\Lambda}_{\ \ \Sigma}=(L^{\Lambda}_{\ \ \Pi})^{-1}
dL^{\Pi}_{\ \ \Sigma}
\end{equation}
 \noindent satisfying the Maurer-Cartan
equation
\begin{equation}
d\Omega^{\Lambda}_{\ \ \Sigma}+\Omega^{\Lambda}_{\ \
\Pi}\land\Omega^{\Pi}_{\ \ \Sigma}=0
\end{equation}
\noindent By appropriate decomposition of the indices, we find:
\begin{eqnarray}
\label{1}&&R^r_{\,\ s}=-P^{r}_{\ \ I}\land P^I_{\ \ s}\\
\label{2}&&R^r_{\,\ 0}=-P^{r}_{\ \ I}\land P^I_{\ \ 0}\\
\label{3}&&R^I_{\,\ J}=-P^I_{\ \ r}\land P^r_{\ \ J}-P^I_{\ \
0}\land P^0_{\ \ J}\\ \label{4}&&\nabla P^I_{\,\ r}=0\\
\label{5}&&\nabla P^I_{\,\ 0}=0
\end{eqnarray}
\noindent where
\begin{eqnarray}
&&R^{rs}\equiv d\Omega^r_{\ \ s}+\Omega^{r}_{\ \
t}\land\Omega^t_{\ \ s}+\Omega^{r}_{\ \ 0}\land\Omega^0_{\ \ s}\\
&&R^{r0}\equiv d\Omega^r_{\ \ 0}+\Omega^{r}_{\ \
t}\land\Omega^t_{\ \ 0}\\ &&R^{IJ}\equiv d\Omega^I_{\ \
J}+\Omega^{I}_{\ \ K}\land\Omega^K_{\ \ J}
\end{eqnarray}
\noindent and we have set
\begin{displaymath}
P^I_{\alpha}=\left\{ \begin{array}{rr}P^I_{\,\ 0}\equiv
\Omega^{I}_{\ \ 0}\\ P^I_{\,\ r}\equiv \Omega^{I}_{\ \
r}\end{array}\right.
\end{displaymath}
\noindent Note that $P^I_0$, $P^I_r$ are the vielbeins of the
coset, while $(\Omega^{rs},\,\ \Omega^{r0})$, $(R^{rs},\,\
R^{ro})$ are respectively the connections and the curvatures of
$SO(4)$ decomposed with respect to the diagonal subgroup
$SU(2)\subset SO(4)$.\\ In terms of the previous definitions, the
ungauged superspace curvatures of the matter coupled theory,
(with $m=0$) are now given by:
{\setlength\arraycolsep{1pt}\begin{eqnarray}
&T^{A}&=\mathcal{D}V^{a}-\frac{i}{2}\ \
\overline{\psi}_{A}\gamma_{a}\psi^A V^{a}=0\\
&R^{ab}&=\mathcal{R}^{ab}\\
&H&=dB+2 e^{-2\sigma}\,\
\overline{\psi}_{A}\gamma_{7}\gamma_{a}\psi^AV^{a}\\
&F^{\Lambda}&=\mathcal{F}^{\Lambda}-ie^{\sigma}L^{\Lambda}_0\epsilon^{AB}
\overline{\psi}_{A}\gamma_{7}\psi_{B}-ie^{\sigma}L^{\Lambda}_r\sigma^{rAB}
\overline{\psi}_{A}\psi_{B}\\
&\rho_{A}&=\mathcal{D}\psi_{A}-\frac{i}{2}\sigma_{rAB}(-\frac{1}{2}\epsilon^{rst}\Omega_{st}-i\gamma_7\Omega_{r0})\psi^B\\
&D\chi_{A}&=\mathcal{D}\chi_{A}-\frac{i}{2}\sigma_{rAB}(-\frac{1}{2}\epsilon^{rst}\Omega_{st}-i\gamma_7\Omega_{r0})\chi^B\\
&R(\sigma)&=d\sigma\\
&\nabla\lambda_{IA}&=\mathcal{D}\lambda_{IA}-\frac{i}{2}\sigma_{rAB}(-\frac{1}{2}\epsilon^{rst}\Omega_{st}-i\gamma_7\Omega_{r0})\lambda_I^B\\
&R^{I}_0(\phi)&\equiv P^{I}_0\\
&R^{I}_r(\phi)&\equiv P^{I}_r
\end{eqnarray}}
\noindent where the last two equations define the "curvatures" of
the matter scalar fields $\phi^i$ as the vielbein of the coset:
\begin{equation}
P^{I}_0\equiv P^{I}_{0 i}d\phi^i\ \ \ \ P^{I}_r\equiv P^{I}_{r
i}d\phi^i
\end{equation}
\noindent where $i$ runs over the $4n$ values of the coset
vielbein world-components.

As in the pure supergravity case one can now write down the
superspace Bianchi identities for the matter coupled curvatures.
The computation is rather long but straightforward. We limit
ourselves to give the new transformation laws of all the physical
fields when matter is present, as derived from the solutions of
the Bianchi identities.
{\setlength\arraycolsep{1pt}\begin{eqnarray} &\delta
V^{a}_{\mu}&=-i\overline{\psi}_{A\mu}\gamma^{a}\varepsilon^A\\
&\delta B_{\mu\nu}&=2 e^{-2\sigma}
\overline{\chi}_{A}\gamma_{7}\gamma_{\mu\nu}\varepsilon^A
-4e^{-2\sigma}\overline{\varepsilon}_A\gamma_7\gamma_{[\mu}\psi_{\nu]}^A\\
 &\delta A^{\Lambda}_{\mu}&=2 e^{\sigma}
 \overline{\varepsilon}^{A}\gamma_{7}\gamma_{\mu}\chi^BL^{\Lambda}_0\epsilon_{AB}+2e^{\sigma}\overline{\varepsilon}^{A}\gamma_{\mu}\chi^{B}L^{\Lambda r}\sigma_{rAB}-e^{\sigma}L_{\Lambda
I}\overline{\varepsilon}^{A}\gamma_{\mu}\lambda^{IB}\epsilon_{AB}+\nonumber\\&&+2ie^{\sigma}L^{\Lambda}_0\overline{\varepsilon}_A\gamma^7\psi_B\epsilon^{AB}+2ie^{\sigma}L^{\Lambda r}\sigma_{r}^{AB}\overline{\varepsilon}_A\psi_B\\
\label{qui}&\delta\psi_{A\mu}&=\mathcal{D}_{\mu}\varepsilon_A+\frac{1}{16}
e^{-\sigma}[T_{[AB]\nu\lambda}\gamma_{7}-T_{(AB)\nu\lambda}](\gamma_{\mu}^{\,\
\nu\lambda}-6\delta_{\mu}^{\nu}\gamma^{\lambda})
\varepsilon^{B}+\nonumber \\
 &&+\frac{i}{32}e^{2\sigma} H_{\nu\lambda\rho}
\gamma_{7}(\gamma_{\mu}^{\,\ \nu\lambda\rho}-3\delta_{\mu}^{\nu}
\gamma^{\lambda\rho})\varepsilon_{A}+\frac{1}{2}\varepsilon_{A}\overline{\chi}^{C}\psi_{C\mu}+\nonumber\\
&&+\frac{1}{2}\gamma_{7}\varepsilon_{A}\overline{\chi}^{C}\gamma^{7}\psi_{C\mu}-\gamma_{\nu}\varepsilon_{A}\overline{\chi}^{C}\gamma^{\nu}\psi_{C\mu}+\gamma_{7}\gamma_{\nu}\varepsilon_{A}\overline{\chi}^{C}\gamma^{7}\gamma^{\nu}\psi_{C\mu}+\nonumber\\
&&-\frac{1}{4}\gamma_{\nu\lambda}\varepsilon_{A}\overline{\chi}^{C}\gamma^{\nu\lambda}\psi_{C\mu}-\frac{1}{4}\gamma_{7}\gamma_{\nu\lambda}\varepsilon_{A}\overline{\chi}^{C}\gamma^{7}\gamma^{\nu\lambda}\psi_{C\mu}\\
\label{quo}&\delta\chi_{A}&=\frac{i}{2}
\gamma^{\mu}\partial_{\mu}\sigma \varepsilon_{A}\!+\!
\frac{i}{16}e^{-\sigma}[T_{[AB]\mu\nu}\gamma_{7}\!+\!T_{(AB)\mu\nu}]\gamma^{\mu\nu}\varepsilon^{B}\!+\!\frac{1}{32}e^{2\sigma}
H_{\mu\nu\lambda}\gamma_{7}\gamma^{\mu\nu\lambda}\varepsilon_{A}\\
&\delta\sigma&=\overline{\chi}_{A}\varepsilon^A\\
\label{qua}&\delta\lambda^{IA}&=-iP^I_{ri}\sigma^{rAB}\partial_{\mu}\phi^{i}\gamma^{\mu}\varepsilon_{B}+iP^I_{0i}\epsilon^{AB}\partial_{\mu}\phi^{i}\gamma^{7}\gamma^{\mu}\varepsilon_{B}+\frac{i}{2}e^{-\sigma}T^{I}_{\mu\nu}\gamma^{\mu\nu}\varepsilon^{A}\\
&P^{I}_{0i}\delta\phi^i&=\frac{1}{2}\overline{\lambda}^{I}_{A}\gamma_{7}\varepsilon^A\\
&P^{I}_{ri}\delta\phi^i&=\frac{1}{2}\overline{\lambda}^{I}_{A}\varepsilon_{B}\sigma_r^{ab}
\end{eqnarray}}
\noindent where we have introduced the "dressed" vector field
strengths
\begin{eqnarray}
&&T_{[AB]\mu\nu}\equiv\epsilon_{AB}L^{-1}_{0\Lambda
}F^{\Lambda}_{\mu\nu}\\
&&T_{(AB)\mu\nu}\equiv\sigma^r_{AB}L^{-1}_{r\Lambda}F^{\Lambda}_{\mu\nu}\\
&&T_{I\mu\nu}\equiv L^{-1}_{I\Lambda }F^{\Lambda}_{\mu\nu}
\end{eqnarray}
\noindent and we have omitted in the transformation laws of the fermions the three-fermions terms of the form $(\chi\chi\varepsilon)$, $(\lambda\lambda\varepsilon)$.\\

\section{The gauging}
The next problem we have to cope with is the gauging of the
matter coupled theory and the determination of the scalar
potential.\\
Let us first consider the ordinary gauging, with $m=0$, which, as
usual, will imply the presence of new terms proportional to the
coupling constants in the supersymmetry transformation laws of
the fermion fields.\\
Our aim is to gauge a compact subgroup of $O(4,n)$. Since in any
case we may gauge only the diagonal subgroup $SU(2)\subset
O(4)\subset H$, the maximal gauging is given by
$SU(2)\otimes\mathcal{G}$ where $\mathcal{G}$ is a $n$-dimensional
subgroup of $O(n)$. According to a well known procedure, we
modify the definition of the left invariant 1-form $L^{-1}dL$ by
replacing the ordinary differential with the
$SU(2)\otimes\mathcal{G}$ covariant differential as follows:
\begin{equation}
\label{nabla}\nabla L^{\Lambda}_{\ \ \Sigma}=d L^{\Lambda}_{\ \
\Sigma}-f_{\Gamma\ \ \Pi}^{\,\ \Lambda} A^{\Gamma} L^{\Pi}_{\ \
\Sigma}
\end{equation}
\noindent where $f^{\Lambda}_{\ \ \Pi\Gamma}$ are the structure
constants of $SU(2)_d\otimes\mathcal{G}$. More explicitly,
denoting with $\epsilon^{rst}$ and $\mathcal{C}^{IJK}$ the
structure constants of the two factors $SU(2)$ and $\mathcal{G}$,
equation (\ref{nabla}) splits as follows:
\begin{eqnarray}
&&\nabla L^{0}_{\ \ \Sigma}=d L^{\Lambda}_{\ \ \Sigma}\\
&&\nabla L^{r}_{\ \ \Sigma}=d L^{r}_{\ \ \Sigma}-g\epsilon^{\,\
r}_{t\
\ s} A^{t} L^{s}_{\ \ \Sigma}\\
&&\nabla L^{I}_{\ \ \Sigma}=d L^{I}_{\ \
\Sigma}-g'\mathcal{C}^{\,\ I}_{K\ \ J} A^{K} L^{J}_{\ \ \Sigma}
\end{eqnarray}
\noindent Setting $\widehat{\Omega}=L^{-1}\nabla L$, one easily
obtains the gauged Maurer-Cartan equations:
\begin{equation}\label{mc}d\widehat{\Omega}^{\Lambda}_{\ \
\Sigma}+\widehat{\Omega}^{\Lambda}_{\ \
\Pi}\land\widehat{\Omega}^{\Pi}_{\ \
\Sigma}=(L^{-1}\mathcal{F}L)^{\Lambda}_{\ \ \Sigma}
\end{equation}
\noindent where $\cF\equiv\cF^{\Lambda}T_{\Lambda}$, $T_{\Lambda}$ being the generators of $SU(2)\otimes\cG$.\\
After gauging, the same decomposition as in eqs. (\ref{1})
-(\ref{5}) now gives:
\begin{eqnarray}
\label{11}&&R^r_{\,\ s}=-P^{r}_{\ \ I}\land P^I_{\ \ s}+(L^{-1}\mathcal{F}L)^{r}_{\ \ s}\\
\label{12}&&R^r_{\,\ 0}=-P^{r}_{\ \ I}\land P^I_{\ \ 0}+(L^{-1}\mathcal{F}L)^{r}_{\ \ 0}\\
\label{13}&&R^I_{\,\ J}=-P^I_{\ \ r}\land P^r_{\ \ J}-P^I_{\ \
0}\land P^0_{\ \ J}+(L^{-1}\mathcal{F}L)^{I}_{\ \ J}\\ \label{14}&&\nabla P^I_{\,\ r}=(L^{-1}\mathcal{F}L)^{I}_{\ \ r}\\
\label{15}&&\nabla P^I_{\,\ 0}=(L^{-1}\mathcal{F}L)^{I}_{\ \ 0}
\end{eqnarray}
Because of the presence of the gauged terms in the coset
curvatures, the new Bianchi Identities are not satisfied by the
old superspace curvatures but we need extra terms in the fermion
field strengths parametrizations, that is, in space-time language,
extra terms in the transformation laws of the fermion fields of
eqs. (\ref{qui}), (\ref{quo}), (\ref{qua}), named ``fermionic
shifts''.
\begin{eqnarray}
\label{del1}&&\delta\psi_{A\mu}=\delta\psi_{A\mu}^{(old)}+S_{AB}(g,g')\gamma_{\mu}\varepsilon^B\\
\label{del2}&&\delta\chi_A=\delta\chi_A^{(old)}+N_{AB}(g,g')\varepsilon^B\\
\label{del3}&&\delta\lambda_A^I=\delta\lambda^{I
(old)}_A+M^I_{AB}(g,g')\varepsilon^B
\end{eqnarray}
\noindent Again, working out the Bianchi identities, one fixes the
explicit form of the fermionic shifts which turn out to be
\begin{eqnarray}
\label{S}&&S_{AB}^{(g,g')}=\frac{i}{24}Ae^{\sigma}\epsilon_{AB}-\frac{i}{8}B_t\gamma^7\sigma^t_{AB}\\
\label{N}&&N_{AB}^{(g,g')}=\frac{1}{24}Ae^{\sigma}\epsilon_{AB}+\frac{1}{8}B_t\gamma^7\sigma^t_{AB}\\
\label{M}&&M^{I(g,g')}_{AB}=(-C^I_t+2i\gamma^7D^I_t)\sigma^t_{AB}
\end{eqnarray}
\noindent where
\begin{eqnarray}\label{AA}&&A=\epsilon^{rst}K_{rst}\\  \label{BB}&&B^i=\epsilon^{ijk}K_{jk0}\\ \label{CC}&&C_I^t=\epsilon^{trs}K_{rIs}\\ \label{DD}&&D_{It}=K_{0It}\end{eqnarray}
\noindent and the threefold completely antisymmetric tensors
$K's$ are the so called "boosted structure constants" given
explicitly by:
\begin{eqnarray}
&&K_{rst}=g\epsilon_{lmn}L^l_{\,\ r}(L^{-1})^{\,\ m}_sL^n_{\,\
t}+g'\mathcal{C}_{IJK}L^I_{\,\ r}(L^{-1})^{\,\ J}_sL^K_{\,\ t}\\
&&K_{rs0}=g\epsilon_{lmn}L^l_{\,\ r}(L^{-1})^{\,\ m}_sL^n_{\,\
0}+g'\mathcal{C}_{IJK}L^I_{\,\ r}(L^{-1})^{\,\ J}_sL^K_{\,\ 0}\\
&&K_{rIt}=g\epsilon_{lmn}L^l_{\,\ r}(L^{-1})^{\,\ m}_IL^n_{\,\
t}+g'\mathcal{C}_{LJK}L^L_{\,\ r}(L^{-1})^{\,\ J}_IL^K_{\,\ t}\\
&&K_{0It}=g\epsilon_{lmn}L^l_{\,\ 0}(L^{-1})^{\,\ m}_IL^n_{\,\
t}+g'\mathcal{C}_{LJK}L^L_{\,\ 0}(L^{-1})^{\,\ J}_IL^K_{\,\ t}
\end{eqnarray}
Actually one easily see that the fermionic shifts (\ref{S})
(\ref{N}) reduce to the pure supergravity $g$ dependent terms of
equations (2.39) and (2.40) of reference \cite{nostro}. (Note
that, since $L^{\Lambda}_{\ \
\Sigma}\rightarrow\delta^{\Lambda}_{\Sigma}$ in absence of
matter, the terms proportional to the Pauli $\sigma$
matrices are simply absent in such a limit.)\\
At this point one could compute the scalar potential of the
matter coupled theory, in terms of the fermionic shifts just
determined, using the well known Ward identity of the scalar
potential which can be derived from the Lagrangian. Since we are
going to perform this derivation once we will introduce also $m$
dependent terms in the fermionic shifts, we just quote, for the
moment the expected result that the potential due only to $g$ and
$g'$ dependent shifts doesn't admit a stable $AdS$ background
configuration. We are thus, led as in the pure supergravity case,
to determine suitable $m$ dependent terms that reduce to the $m$
terms of eqs. (2.56) and (2.57) of reference \cite{nostro} in
absence of matter multiplets (one can see that the simple-minded
ansatz of keeping exactly the same form for the $m$ dependent
terms as in the pure Supergravity case is not consistent with the
gauged superspace
Bianchi identities).\\
It turns out that a consistent solution for the relevant $m$ terms
to be added to the fermionic shifts, implies the presence of the
coset representatives; that is, the $m$-terms must also be
"dressed" with matter scalar fields as it happens for the $g$ and
$g'$ dependent terms. Explicitly, the Bianchi identities solution
for the new fermionic shifts is:
\begin{eqnarray}
\label{mS}&&
S_{AB}^{(g,g',m)}=\frac{i}{24}[Ae^{\sigma}+6me^{-3\sigma}(L^{-1})_{00})\epsilon_{AB}-\frac{i}{8}[B_te^{\sigma}-2me^{-3\sigma}(L^{-1})_{i0}]\gamma^7\sigma^t_{AB}\\
\label{mN}&&N_{AB}^{(g,g',m)}\!\!=\!\!\frac{1}{24}[Ae^{\sigma}-18me^{-3\sigma}(L^{-1})_{00})]\epsilon_{AB}+\frac{1}{8}[B_te^{\sigma}+6me^{-3\sigma}(L^{-1})_{i0}]\gamma^7\sigma^t_{AB}\\
\label{mM}&&M^{I(g,g',m)}_{AB}=(-C^I_t+2i\gamma^7D^I_t)e^{\sigma}\sigma^t_{AB}-2me^{-3\sigma}(L^{-1})^I_{\
\ 0}\epsilon_{AB}
\end{eqnarray}

\section{The scalar potential}
The simplest way to derive the scalar potential is to use the
supersymmetry Ward identity which relates the scalar potential to
the fermionic shifts in the transformation laws \cite{fema}. In
order to retrieve such identity it is necessary to have the
relevant terms of the Lagrangian of the gauged theory. These
terms are actually the kinetic ones and the "mass" terms given in
the following equation:
\begin{eqnarray}
\label{lag}&&(detV)^{-1}\mathcal{L}=-\frac{1}{4}\mathcal{R}
-\frac{1}{8}e^{2\sigma}\mathcal{N}_{\Lambda\Sigma}\widehat{\mathcal{F}}^{\Lambda}_{\mu\nu}\widehat{\mathcal{F}}^{\Sigma\mu\nu}+\partial^{\mu}\sigma\partial_{\mu}\sigma
-\frac{1}{4}(P^{I0}_{\mu}P^{\mu}_{I0}
+P^{Ir}_{\mu}P^{\mu}_{Ir})+\nonumber\\
&&-\frac{i}{2}\overline{\psi}_{A\mu}\gamma^{\mu\nu\rho}D_{\nu}\psi^{A}_{\rho}+\frac{i}{8}\overline{\lambda}^I_A\gamma^{\mu}D_{\mu}\lambda^A_I-2i\overline{\chi}_A\gamma^{\mu}D_{\mu}\chi^A
+2i\overline{\psi}_{\mu}^A\gamma^{\mu\nu}\overline{S}_{AB}\psi_{\nu}^B+\nonumber\\
&&+4i\overline{\psi}_{\mu}^A\gamma^{\mu}\overline{N}_{AB}\chi^{B}+\frac{i}{4}\overline{\psi}_{\mu}^A\gamma^{\mu}\overline{M}_{AB}^I\lambda^{B}_I+\mathcal{W}(\sigma\phi^i;g,g',m)+\dots
\end{eqnarray}
\noindent where
\begin{equation}
\label{kin}\mathcal{N}_{\Lambda\Sigma}=L_{\Lambda}^{\,\
0}L^{-1}_{0\Sigma}+L_{\Lambda}^{\,\
i}L^{-1}_{i\Sigma}-L_{\Lambda}^{\,\ I}L^{-1}_{I\Sigma}
\end{equation}
\noindent is the vector kinetic matrix,
$\widehat{\mathcal{F}}^{\Lambda}_{\mu\nu}\equiv\mathcal{F}^{\Lambda}_{\mu\nu}-m\delta^{\Lambda}_0B_{\mu\nu}$ and $\cW$ is minus the scalar potential.\\
 In equation
(\ref{lag}) there appear ``barred mass-matrices''
$\overline{S}_{AB},\,\ \overline{N}_{AB},\,\ \overline{M}^I_{AB}$
which are slightly different from the fermionic shifts defined in
eqs. (\ref{mS}), (\ref{mN}), (\ref{mM}). Actually they are
defined by:
\begin{equation}
\label{pippo}\overline{S}_{AB}=-S_{BA},\ \ \ \
\overline{N}_{AB}=-N_{BA},\ \ \ \ \overline{M}^I_{AB}=M^I_{BA}
\end{equation}
\noindent Definitions (\ref{pippo}) stem from the fact that the
shifts defined in eqs. (\ref{mS}), (\ref{mN}), (\ref{mM}) are
matrices in the eight-dimensional spinor space, since they
contain the $\gamma_7$ matrix; as will be seen in a moment, such
definition is actually necessary in order to satisfy the
supersymmetry Ward identity.\\
 Indeed, let us perform the
supersymmetry variation of (\ref{lag}), keeping only the terms
proportional to $g$, $g'$ or $m$, and to the current
$\overline{\psi}_{A\mu}\gamma^{\mu}\epsilon^A$; we find the
following Ward identity, :
\begin{equation}
\label{ward}\delta^C_A\mathcal{W}=20\overline{S}^{AB}S_{BC}+4\overline{N}^{AB}N_{BC}+\frac{1}{4}\overline{M}^{AB}_IM^I_{BC}
\end{equation}
\noindent However we note that, performing the supersymmetry
variation, the gauge terms also give  rise to extra terms
proportional to the current
$\overline{\psi}_{A\mu}\gamma^7\gamma^{\mu}\epsilon^A$, which
have no counterpart in the term containing the potential $\cW$.
Because of the definition of the barred mass matrices in eq.
(\ref{pippo}) it is easily seen that such "$\gamma^7$-terms",
arising from $\overline{S}^{AB}S_{BC}$ and
$\overline{N}^{AB}N_{BC}$ cancel
against each other.\\
As far as the term $\overline{M}^{AB}_IM^I_{BC}$ is concerned,
 the same mechanism of cancellation again applies to the terms
proportional to
$\overline{\psi}_{A\mu}\gamma_7\gamma^{\mu}\epsilon^A\sigma^{rA}_C$;
there is, however, a residual dangerous term of the form
\begin{equation}
\delta^A_C\overline{\psi}_{A\mu}\gamma^{\mu}\gamma^7D^I_{\,\
s}C_I^{\,\ s}\epsilon^C
\end{equation}
\noindent One can show that this term vanishes identically owing
to the non trivial relation
\begin{equation}
\label{abigaille}D^I_tC_I^t=0
\end{equation}
\noindent Equation (\ref{abigaille}) can be shown to hold  using
the pseudo-orthogonality relation $L^T\eta L=\eta$ among the coset
representatives and the Jacobi identities $C_{I[JK}C_{L]MN}=0$,
$\epsilon_{r[st}\epsilon_{l]mn}=0$. This is a non trivial check of our computation.\\
It now follows that the Ward identity eq. (\ref{ward}) is indeed
satisfied since all the terms on the r.h.s., once the "$\gamma^7$-terms" have been cancelled,  are proportional to $\delta^C_A$.\\
Using the expressions (\ref{mS}), (\ref{mN}), (\ref{mM}),
(\ref{pippo}) in equation (\ref{ward}), we obtain the explicit
form of the scalar potential
{\setlength\arraycolsep{1pt}\begin{eqnarray}\label{pot}\mathcal{W}(\phi)=&&
5\,\ \{
[\frac{1}{12}(Ae^{\sigma}+6me^{-3\sigma}L_{00})]^2+[\frac{1}{4}(e^{\sigma}B_i-2me^{-3\sigma}L_{0i})]^2\}+\nonumber\\
\nonumber\\ &&- \{
[\frac{1}{12}(Ae^{\sigma}-18me^{-3\sigma}L_{00})]^2+[\frac{1}{4}(e^{\sigma}B_i+6me^{-3\sigma}L_{0i})]^2\}+\nonumber\\
&&-\frac{1}{4}\{C^I_{\,\ t}C_{It}+4D^I_{\,\ t}D_{It}\}\,\
e^{2\sigma}-m^2e^{-6\sigma}L_{0I}L^{0I}
\end{eqnarray}}

\noindent Expanding the squares in equation (\ref{pot})the
potential $\mathcal W$can be alternatively written as follows:
\begin{eqnarray}
&&\mathcal{W}=e^{2\sigma}[\frac{1}{36}A^2+\frac{1}{4}B^iB_i-\frac{1}{4}(C^I_{\,\
t}C_{It}+4D^I_{\,\
t}D_{It})]-m^2e^{-6\sigma}\mathcal{N}_{00}+\nonumber\\
&&+me^{-2\sigma}[\frac{2}{3}AL_{00}-2B^iL_{0i}]
\end{eqnarray}
\noindent where $\mathcal{N}_{00}$ is the 00 component of the
vector kinetic matrix defined in eq. (\ref{kin}).\\
We now show that, apart from other possible extrema not
considered here, a stable supersymmetric extremum of the potential
$\mathcal{W}$ is found to be the same as in the case of pure
supergravity, that is we get an $AdS$ supersymmetric background
only for $g=3m$. .\\
A further issue related to the scalar potential, which is an
important check of all our calculation, is the possibility of
computing the masses of the scalar fields by varying the
linearized kinetic terms and the potential of (\ref{lag}), after
power expansion of $\mathcal{W}$ up to the second order in the
scalar fields $q^I_{\alpha}$. We find:
\begin{eqnarray}
&&(\frac{\partial^2\mathcal{W}}{\partial\sigma^2})_{\sigma=q=0,
g=3m}=48m^2\\
&&(\frac{\partial^2\mathcal{W}}{\partial q^{I0}\partial q^{J0}}
)_{\sigma=q=0,
g=3m}=8m^2\delta^{IJ}\\
&&(\frac{\partial^2\mathcal{W}}{\partial q^{Ir}\partial q^{Js}
})_{\sigma=q=0, g=3m}=24m^2\delta^{IJ}\delta^{rs}
\end{eqnarray}
\noindent The linearized equations of motion become:
\begin{eqnarray}
&&\Box\sigma-24m^2\sigma=0\\
&&\Box q^{I0}-16m^2 q^{I0}=0\\
&&\Box q^{Ir}-24m^2 q^{Ir}=0
\end{eqnarray}
\noindent If we use as mass unity the
\def\IP{\relax{\rm I\kern-.18em P}} inverse $AdS$ radius, which
in our conventions is $R^{-2}_{AdS}=4m^2$ we get:
\begin{eqnarray}
&&m^2_{\sigma}=-6\nonumber\\
&&m^2_{q^{I0}}=-4\nonumber\\
\label{massa}&&m^2_{q^{Ir}}=-6
\end{eqnarray}
\noindent These values should be compared with the results
obtained in reference \cite{fkpz} where the supergravity and
matter multiplets of the $AdS_6\,\ F(4)$ theory were constructed
in terms of the singleton fields of the 5-dimensional conformal
field theory, the singleton being given by hypermultiplets
transforming in the fundamental of $\mathcal{G}\equiv E_7$. It is
amusing to see that the values of the masses of the scalars
computed in terms of the conformal dimensions are exactly the
same as those given in equation (\ref{massa}).\\ This coincidence
can be considered as a non trivial check of the $AdS/CFT$
correspondence in six versus five dimensions.\\
To make contact with what follows we observe that the scalar
squares masses in $AdS_{d+1}$ are given by the $SO(2,d)$
quadratic Casimir \cite{flafro} \be m^2=E_0(E_0-d) \ee \noindent
They are negative in the interval $\frac{d-2}{2}\leq E_0<d$ (the
lower bound corresponding to the unitarity bound i.e. the
singleton) and attain the Breitenlohner-Freedman bound
\cite{breit} when $E_0=d-E_0$ i.e. at $E_0=\frac{d}{2}$ for which
$m^2=-\frac{d^2}{4}$. Conformal propagation correspond to
$m^2=-\frac{d^2-1}{4}$ i.e. $E_0=\frac{d\pm 1}{2}$. This is the
case of the dilaton and triplet matter scalars.

\section{$F(4)\otimes\cG$ Superconformal Field Theory}
Here we describe the basics of the  $F(4)$ highest weight unitary
irreducible representations ``UIR's'' and exhibit two towers of
short representations which are relevant for a K-K analysis of
type IIA theory on  (warped) $AdS_6\otimes S^4$
\cite{oz},\cite{clp}.\\ We will not consider here the $\cG$
representation properties but we will only concentrate on the
supersymmetric structure.\\ Recalling that the even part of the
$F(4)$ superalgebra is $SO(2,5)\otimes SU(2)$, from a general
result on Harish-Chandra modules \cite{f},\cite{min}, \cite{g},
\cite{h} of $SO(2,2n+1)$ we know that there are only a spin 0 and
a spin 1/2 singleton unitary irreducible representations
\cite{flafro}, which, therefore, merge into a unique
supersingleton representation of the $F(4)$ superalgebra: the
hypermultiplet \cite{fkpz}.\\ To describe shortening is useful to
use a harmonic superfield language \cite{fanta1}.\\ The harmonic
space is in this case the 2-sphere\footnote{The sphere is the
simplest example  of ``flag manifold'' whose geometric structure
underlies the construction of harmonic superspaces \cite{hh}}
$SU(2)/U(1)$, as in $N=2,\,\ d=4$ and $N=1,\,\ d=6$. A highest
weight UIR of $SO(2,5)$ is determined by $E_0$ and a UIR of
$SO(5)\simeq Usp(4)$, with Dynkin labels $(a_1,\,\ a_2)$
\footnote {Note that the $Usp(4)$ Young labels $h_1,h_2$ are
related to $a_1,a_2$ by $a_1=2h_2; a_2=h_1-h_2$.}. We will denote
such representations by $\cD(E_0,\,\ a_1,\,\ a_2)$. The two
singletons correspond to $E_0=3/2$, $a_1=a_2=0$ and $E_0=2$,
$a_1=1$, $a_2=0$.\\ In the $AdS/CFT$ correspondence $(E_0,\,\
a_1,\,\ a_2)$ become the conformal
dimension and the Dynkin labels of $SO(1,4)\simeq Usp(2,2)$.\\
The highest weight UIR of the $F(4)$ superalgebra will be denoted
by $\cD(E_0,\,\ a_1,\,\ a_2;\,\ I)$ where $I$ is the $SU(2)$
$R$-symmetry quantum number (integer or half integer).\\ We will
show shortly that there are two (isolated) series of UIR's which
correspond respectively to $1/2$ BPS short multiplets (analytic
superfields) and intermediate short superfields. The former have
the property that they form a ring under multiplication, as the
chiral fields in $d=4$ \cite{fesoca}.\\ The first series is the
massive
 tower of short vector multiplets  whose lowest
members is a massless vector multiplet in $Adj\cG$ corresponding
to the conserved currents of the  $\cG$ global symmetry of the
five dimensional
conformal field theory.\\
The other series is the tower of massive graviton multiplets,
which exhibit "intermediate shortening" and it is not of BPS
type. Its lowest member is the supergravity multiplet which
contains the $SU(2)$ $R$-symmetry current and the stress-tensor
among the  superfield components.

\subsection{$F(4)$ superfields}

The basic superfield is the supersingleton hypermultiplet
$W^A(x,\theta)$, which satisfies the constraint
\begin{equation}\label{hyper}
D_{\alpha}^{(A}W^{B)}(x,\theta)=0
\end{equation}
\noindent corresponding to the irrep. $\cD(E_0=\frac{3}{2},0,0;I=\frac{1}{2})$ \cite{fanta1}.\\
\noindent By using harmonic superspace, $(x,\,\ \theta_I,\,\
u^I_i)$, where $\theta_I=\theta_iu^i_I$, $u^i_I$ is the coset
representative of $SU(2)/U(1)$ and $I$ is the charge $U(1)$-label,
from  the covariant derivative algebra \be
\{D^A_{\alpha},D^{jB}_{\beta}\}=i\epsilon^{AB}\partial_{\alpha\beta}
\ee
 \noindent we have
\be \{D^I_{\alpha},D^I_{\beta}\}=0 \ \ \ \
D^I_{\alpha}=D^i_{\alpha}u^I_i \ee \noindent Therefore from eq.
(\ref{hyper}) it follows the $G$-analytic constraint: \be
D_{\alpha}^{1}W^{1}=0\ee \noindent which implies \be
W^1(x,\theta)=\varphi^1+\theta^{\alpha}_2\zeta_{\alpha}+d.t. \ee
 \noindent (d.t. means ``derivative terms'').\\
Note that $W^1$ also satisfies \be D^2_{\alpha}D^{2\alpha}W^1=0
\ee
 \noindent because there is no such scalar component\footnote{This is rather similar to the treatment of the (1,0) hypermultiplet in $D=6$ \cite{fanta2}} in $W^1$.\\
$W^1$ is a Grassman  analytic  superfield, which is also harmonic
(that is ${\bf D}^1_2W^1=0$ where, using notations of reference
\cite{fesoca}, ${\bf D}^1_2$ is the step-up operator of the
$SU(2)$ algebra acting on  harmonic superspace).\\ Since $W^1$
satifies $D^1W^1=0$, any $p$-order polynomial \be
\label{wp}I_p(W^1)=(W^1)^p\ee \noindent will also have the same
property, so these operators form a ring under multiplication
\cite{fesoca}, they are the 1/2 BPS states of the $F(4)$
superalgebra and represent massive vector
multiplets $(p>2)$, and massless bulk gauge fields for $p=2$.\\
The above multiplets correspond to the $D(E_0=3I,0,0;I=\frac{p}{2})$ h.w. U.I.R.'s of the $F(4)$ superalgebra.\\
Note also that if $W^1$ carries a pseudo-real representation of
the flavor group $\cG$ (e.g. {\bf 56} of $\cG=E_7$) then $W^1$
satisfies a reality condition \be (W^1)^*=W^2 \ee
 \noindent corresponding to the superfield constraint
\be
(W^A)^{*\Lambda}=\epsilon_{AB}\Omega_{\Lambda\Sigma}W^{jB\Sigma}
\ee
 \noindent The $SU(2)$ quantum numbers of the $W^{1p}$ superfield
 components are:
\begin{eqnarray*} &&(\theta)^0\hspace{19 mm}spin\,\ 0\hspace{20
mm}I=\frac{p}{2}\\ &&(\theta)^1\hspace{19
mm}spin\frac{1}{2}\hspace{20 mm}I=\frac{p}{2}-\frac{1}{2}\\
&&(\theta)^2\hspace{10 mm}spin\,\ 0 - spin\,\ 1\hspace{10
mm}I=\frac{p}{2}-1\\ &&(\theta)^3\hspace{19
mm}spin\frac{1}{2}\hspace{20 mm}I=\frac{p}{2}-\frac{3}{2}\\
&&(\theta)^4\hspace{19 mm}spin\,\ 0\hspace{20 mm}I=\frac{p}{2}-2
\end{eqnarray*}
\noindent Note that the $(\theta)^4$ component is missing for
$p=2$, $p=3$, while the $(\theta)^3$ component is missing for
$p=2$.
However the total number of states is $8(p-1)$ both for boson and fermion fields ($p\geq 2$).\\
 The AdS squared mass for scalars is
\be m_s^2=E_0(E_0-5) \ee
 \noindent so there are three families of scalar states with
\begin{eqnarray*}
&&m^2_1=\frac{3}{4}p(3p-10)\hspace{20 mm}p\geq 2\\
&&m^2_2=\frac{1}{4}(3p+2)(3p-8)\hspace{10 mm}p\geq 2\\
&&m^2_3=\frac{1}{4}(3p+4)(3p-6)\hspace{10 mm}p\geq 4\\
\end{eqnarray*}
\noindent The only scalars states with $m^2<0$ are the scalar in
the massless vector multiplet $(p=2)$ with $m^2_1=-6$, $m^2_2=-4$
(no states with $m^2=0$ exist) and in the $p=3$ multiplet with
$m^2=-\frac{9}{4}$.\\ We now consider the second "short" tower
containing the graviton supermultiplet and its recurrences.\\ The
graviton multiplet is given by $W^1\overline{W}^1$. Note that
such superfield is not $G$-analytic, but it satisfies \be
D^1_{\alpha}D^{1\alpha}(W^1\overline{W}^1)=D^2_{\alpha}D^{2\alpha}(W^1\overline{W}^1)=0
\ee
 \noindent this multiplet is the $F(4)$ supergravity multiplet.
 Its lowest component, corresponding to  the dilaton in $AdS_6$ supergravity multiplet,
 is a scalar with $E_0=3$ $(m^2=-6)$ and $I=0$.\\
 The tower is obtained as follows
\be \label{magra}G_{q+2}(W)=W^1\overline{W}^1(W^1)^q \ee
 \noindent where the massive graviton, described in eq..
 (\ref{magra}) has $E_0=5+\frac{3}{2}q$ and $I=\frac{q}{2}$.\\
 Note that the $G_{q+2}$ polynomial, although not $G$-analytic,
 satisfies the constraint
\be D^1_{\alpha}D^{1\alpha}G_{q+2}(W)=0 \ee
 \noindent so that it corresponds to a short representation with
 quantized dimensions and highest weight given by
 $D(E_0=3+3I,0,0;I=\frac{q}{2})$.\\
 We call these multiplets, following \cite{fanta2}, "intermediate
 short" because, although they have some missing states, they are
 not BPS in the sense of supersymmetry. In fact they do not form a
 ring under multiplication.\\
It is worthwhile to mention that the towers given by (\ref{wp}),
(\ref{magra}) correspond to the two
 isolated series of UIR's of the $F(4)$ superalgebra argued to exist in
 \cite{min}.\\
There are also long spin 2 multiplets containing $2^8$ state where $E_0$ is not quantized and satisfies the bound $E_0\geq 6$.\\
  Finally let us make some comments on the role played by the flavour symmetry $\cG$.\\
It is clear that,
 since the supersingleton $W^1$ is in a representation of $\cG$ (other than the gauge group of the world-volume theory),
 the $I_p$ and $G_{q+2}$ polynomials will appear in the $p$-fold and
 $(q+2)$-fold tensor product representations of the $\cG$ group.
 This representation is in general reducible, however the 1/2 BPS
 states must have a first component totally symmetric in the $SU(2)$ indices  and, therefore, only
 certains $\cG$ representations survive.\\ Moreover in the $(W^1)^2$ multiplet,
 corresponding to the massless $\cG$- gauge vector multiplets in $AdS_6$, we must pick up the adjoint  representation
 $Adj\cG$ and in $W^1\overline{W}^1$, corresponding to the graviton multiplet, we must pick up  the $\cG$  singlet representation.\\
 However in principle there can be representations in the higher
 symmetric and antisymmetric products, and the conformal field theory should
  tell us which products remains, since the flavor symmetry
  depends on the specific dynamical model.\\

  The states discussed in this paper are expected to appear \cite{oz}, \cite{clp} in the
  K-K analysis of IIA massive supergravity on warped $AdS_6\otimes S^4$. It is amusing that superconformal
 field theory largely predicts the spectrum just from symmetry cosiderations.
  What is new in the $F(4)$ theory is the fact that, since it is
  not a theory with maximal symmetry, it allows in principle some
  rich dynamics and more classes of short representations than the
  usual compactification on spheres.\\
The K-K reduction is related to the horizon geometry of the $D4$ branes in a $D8$ brane background in presence of $D0$ branes \cite{fkpz}, \cite{oz}.\\
Conformal theories at fixed points of $5d$ gauge theories exist
\cite{smi} which exibit global symmetries $E_{N_f+1}\supset
SO(2N_f)\otimes U(1)$,  where $N_f\geq 1$ is the number of flavors
 ($N_f$ $D8$ branes) and $U(1)$ is the ``instantons charge'' (dual to the $D0$ brane charge).\\
The $E$ exceptional series therefore unifies perturbative and non perturbative series of the gauge theory.\\
It is natural to conjecture that a conformal fixed point $5d$
theory  can be described by a singleton supermultiplet
 in the fundamental rep. of $E_{N_f+1}$. For the exceptional groups $N_f\geq 5$ these are the {\bf 27} of $E_6$,
the {\bf 56} of $E_7$ and the {\bf 248} of $E_8$ which are
respectively complex, pseudo-real and real. The $E_7$ case was
considered in ref. \cite{fkpz}. States coming from wrapped $D8$
branes will carry a non trivial representation of $SO(2N_f)$,
which, together with some other states, must complete
representations of $E_{N_f+1}$. It is possible that from the
knowledge of $SO(2N_f)$ quantum numbers of supergravity in
$D4-D8$ background one may infer the spectrum of $E_{N_f+1}$
representations and then to realize these states in terms of
boundary composite conformal operators.We hope to return to the
above issues in a future investigation.

+


\begin{thebibliography}{100}
\bibitem{nostro} R. D'Auria, S.Ferrara and S.Vaul\'a,
hep-th/0006107
\bibitem{rom} L.J. Romans, Nucl. Phys. {\bf B269} (1986) 691
\bibitem{nahm} M. Scheunert, W. Nahm and V. Rittenberg, J. Math. Phys. {\bf 17} (1976) 1640; W. Nahm, Nucl. Phys. {\bf B135} (1978) 149
\bibitem{bagu} I. Bars and M. G\"{u}naydin, Comm. Math. Phys. {\bf 91} (1983) 31
\bibitem{malda} Juan M. Maldacena, Adv.Theor.Math.Phys. {\bf 2} (1998) 231, Int.J.Theor.Phys. {\bf 38}
(1999) 1113, hep-th/9711200; S. S. Gubser, I. R. Klebanov, A. M.
Polyakov, Phys.Lett. {\bf B428} (1998) 105, hep-th/9802109;  E.
Witten, Adv.Theor.Math.Phys. {\bf 2} (1998) 253, hep-th/9802150
\bibitem{rass} O. Aharony, S. S. Gubser, J. Maldacena, H. Ooguri and Y.
Oz, Phys.Rept. {\bf 323} (2000) 183, hep-th/9905111

\bibitem{fkpz} S. Ferrara, A. Kehagias, H. Partouche, A.
Zaffaroni, Phys.Lett. {\bf B431} (1998) 57, hep-th/9804006
\bibitem{oz} A. Brandhuber and Y. Oz,  Phys.Lett. {\bf B460} (1999) 307
\bibitem{clp} M. Cveti\v{c}, H. L\"{u} and C.N. Pope, Phys.Rev.Lett. {\bf 83} (1999)
5226, hep-th/9906221
\bibitem{smi} N. Seiberg, Phys. Lett. {\bf B388} (1966) 753,
hep-th/960811; D. R. Morrison and N. Seiberg, Nucl. Phys. {\bf
B483} (1997) 229, hep-th/9609071; P. Intriligator, D. R. Morrison
and N. Seiberg, Nucl. Phys. {\bf B497} (1997) 56, hep-th/9702198
\bibitem{min} S. M. Minwalla, Theor. Math. Phys. {\bf 2} (1998)
781
\bibitem{fanta1} A. Galperin, E. Ivanov, S. Kalitzin, V.
Ogievetsky and E. Sokatchev, Class. Quant. Grav. {\bf 1} (1984)
469
\bibitem{bible} L. Castellani, R. D'Auria and P. Fr\'e,
Supergravity and Superstrings Vol 2 pag 794, World Scientific
(1991)
\bibitem{anna} A. Ceresole and G.G. Dall'Agata, hep-th/0004111
\bibitem{breit} P. Breitenlohner and D. Z. Freedman, Phis. Lett. {\bf B115} (1982) 197; Ann. of Phys. {\bf 144} (1982) 249; G. W. Gibbons, C. M. Hull and N. P. Warner, Nucl. Phys. {\bf B218} (1983) 173; W. Boucher, Nucl. Phys. {\bf B242} (1984) 282
 \bibitem{fema} S. Ferrara and L. Maiani, Proc. V Silarg Symposium (World Scientific, Singapore) (1986); S. Cecotti, L. Girardello and M. Porrati, Nucl. Phys. {\bf B268}(1986) 295; A. Ceresole, R. D'Auria, S. Ferrara, P. Fr\'e and E. Maina, Phys. Lett. {\bf B268} (1986) 317
\bibitem{f} W. Siegel, Int. Jou. Math. Phys. {\bf A4} (1989) 2015
\bibitem{g} E. Angelopoulos and M. Laoues, Rev. Math. Phys. {\bf 10}
(1998) 271, hep-th/9806100
\bibitem{h} S. Ferrara, C. Fronsdal, hep-th/0006009
\bibitem{flafro} M. Flato and G. Fronsdal, Lett. Math. Phys. {\bf 2} (1978) 421; Phys. Lett. {\bf 97B} (1980) 236; J. Math. Phys. {\bf 22} (1981) 1100; Phys. Lett. {\bf B172} (1986) 412
\bibitem{fesoca} S. Ferrara and E. Sokatchev, hep-th/000515
\bibitem{hh} G. G. Hartwell and P. S. Howe, Class. Quant. Grav. {\bf 12} (1995) 1823; Int. J. Mod. Phys. {\bf 10} (1995) 3901
\bibitem{fanta2} S. Ferrara and E. Sokatchev, hep-th/0001178
\end{thebibliography}
\end{document}